\documentclass[journal,onecolumn, draftcls, 12pt]{IEEEtran} % for arxiv

\usepackage[utf8]{inputenc}
\usepackage[T1]{fontenc}
\usepackage[hidelinks]{hyperref}
\usepackage{url}
\usepackage{ifthen}
\usepackage{cite}
\usepackage[cmex10]{amsmath} % Use the [cmex10] option to ensure complicance
\usepackage{amsfonts,amssymb}
\usepackage{color}
\usepackage{ifthen}
\usepackage[arxiv]{optional} % 2col/arxiv

\begin{document}
\title{Sequential decomposition of discrete memoryless channel with noisy feedback}

 %%% Single author, or several authors with same affiliation:
 \author{%
   \IEEEauthorblockN{Deepanshu Vasal}
   %\IEEEauthorblockA{University of Michigan\\
     %                Ann Arbor, MI 48109, USA\\
     %                Email: \texttt{\{anastas,pradhanv\}@umich.edu}}
 }

\maketitle

%%%%%%
%% Abstract:
%% If your paper is eligible for the student paper award, please add
%% the comment "THIS PAPER IS ELIGIBLE FOR THE STUDENT PAPER
%% AWARD." as a first line in the abstract.
%% For the final version of the accepted paper, please do not forget
%% to remove this comment!
%%
\begin{abstract}
In this paper, we consider a discrete memoryless point to point channel with noisy feedback, where there is a sender with a private message that she wants to communicate to a receiver by sequentially transmitting symbols over a noisy channel. After each transmission, she receives a noisy feedback of the symbol received by the receiver. The goal is to design transmission control strategy of the sender that minimize the average probability of error. This is an instance of decentralized control of information where the two controllers, the sender and the receiver have no common information. There exist no methodology in the literature that provides a notion of ``state" and a dynamic program to find optimal policies for this problem.
In this paper, we construct state of the system, based on which we provide a sequential decomposition methodology that finds optimum policies within the class of Markov strategies with respect to this state (which are not necessarily globally optimum). This allows to decompose the problem across time and reduce the complexity dependence on time from double exponential to linear in time.

\end{abstract}

%%%%%%%%%%%%%%%%%%%%%%%%%%%
%%%%%%% DEFINITIONS %%%%%%%
%%%%%%%%%%%%%%%%%%%%%%%%%%%

\def\cE{\mathcal{E}}
\def\cX{\mathcal{X}}
\def\cY{\mathcal{Y}}
\def\cZ{\mathcal{Z}}
\def\cW{\mathcal{W}}
\def\cP{\mathcal{P}}
\def\cU{\mathcal{U}}
\def\cV{\mathcal{V}}
\def\cR{\mathcal{R}}
\def\cC{\mathcal{C}}
\def\cS{\mathcal{S}}
\def\cF{\mathcal{F}}
\def\cG{\mathcal{G}}
\def\cB{\mathcal{B}}

\def\tw{\tilde{w}}
\newcommand{\eq}[1]{\begin{align}#1\end{align}}
\newcommand{\seq}[1]{\begin{subequations}#1\end{subequations}}
\newcommand{\lb}[1]{\left\{ \begin{array}{ll} #1 \end{array} \right.}
\newcommand*\dif{\mathop{}\!\mathrm{d}}
\newcommand{\bm}[1]{\begin{bmatrix}#1\end{bmatrix}}
\newcommand{\bit}[1]{\begin{itemize}#1\end{itemize}}

\newcommand{\cA}{\mathcal{A}}

\newcommand{\cH}{\mathcal{H}}

\newcommand{\tcC}{\tilde{\mathcal{C}}}
\newcommand{\hV}{\hat{V}}
\newcommand{\hu}{\hat{u}}
\newcommand{\hN}{\hat{N}}
\newcommand{\tx}{\tilde{x}}
\newcommand{\hz}{\hat{z}}
\newcommand{\hw}{\hat{w}}
\newcommand{\ha}{\hat{a}}
\newcommand{\hx}{\hat{x}}
\newcommand{\hcC}{\hat{\mathcal{C}}}
\newcommand{\hD}{\hat{D}}
\newcommand{\cN}{\mathcal{N}}
\newcommand{\tgamma}{\tilde{\phi}}
\newcommand{\Qx}[1]{Q^{i}(#1)}
\newcommand{\Qw}[1]{Q_w^{i}(#1)}
\newcommand{\defeq}{\buildrel\triangle\over =}
\newcommand{\apos}{\textsc{\char13}}
\newcommand{\pushright}[1]{\ifmeasuring@ #1 \else\omit\hfill$\displaystyle#1$\fi\ignorespaces}
\newcommand{\pushleft}[1]{\ifmeasuring@ #1 \else\omit$\displaystyle#1$\hfill\fi\ignorespaces}
\newcommand{\nn}{\nonumber}
\newcommand{\rd}{\right.}
\newcommand{\ld}{\left.}

\newtheorem{lemma}{Lemma}
\newtheorem{fact}{Fact}
\newtheorem{theorem}{Theorem}

\newcommand{\ve}[1]{\underline{#1}}
\newcommand{\eqdef}{\stackrel{\scriptscriptstyle \triangle}{=}}
\newcommand{\mdef}{\stackrel{\text{\tiny def}}{=}}
\newcommand{\E}{\mathbb{E}}
\def\Real{\mathbb{R}}
\def\P{\mathbb{P}}

%%%%%%%%%%%%%%%%%%%%%%%%%%%%%%%%%%%%%%%%%%%%%%%%%%%%%%%%%%%%%%%%%%%%%%%%%%%%%%%%%
%%%%%%%%%%%%%%%%%%%%%%%%%%%%%%%%%%%%%%%%%%%%%%%%%%%%%%%%%%%%%%%%%%%%%%%%%%%%%%%%%
%%%%%%%%%%%%%%%%%%%%%%%%%%%%%%%%%%%%%%%%%%%%%%%%%%%%%%%%%%%%%%%%%%%%%%%%%%%%%%%%%
%%%%%%%%%%%%%%%%%%%%%%%%%%%%%%%%%%%%%%%%%%%%%%%%%%%%%%%%%%%%%%%%%%%%%%%%%%%%%%%%%
\vspace*{-0.15cm}
\section{Introduction}
Communication is ubiquitous in our lives, be it wireless devices such as cellphones, IoT devices, satellite communication, or wired communication such as Ethernet and more. The problem of communicating a message reliably and efficiently with least delay and energy is a fundamental problem which has been attempted to be addressed from many different tools in engineering. The fundamentals of digital communication was laid down by Shannon in his pioneering work in~\cite{Sh48}. Since then there has been significant effort on finding coding schemes that minimize probability of error and achieve capacity. There has been a lot of focus on a point to point discrete memoryless channel for which efficient codes such as Turbo codes, LDPC codes etc have been formulated. For a point to point channel with feedback, it is known that the feedback doesn't increase the capacity~\cite{Sh56} but it can significantly improve the error exponents from exponential to double exponential~\cite{ScKa66}.

There have been multiple works proposing transmission schemes for several instances of channel with \emph{noiseless} feedback such as Horstein's scheme~\cite{Ho63} for binary symmetric channel (BSC), and Schalkwijk and Kailath~\cite{ScKa66} for an additive white Gaussian noise (AWGN) channel, all of which were generalized by a posterior matching scheme (PMS)~\cite{ShFe08} for an arbitrary channel. However, it is known that these schemes perform rather poorly when the feedback is even slightly noisy~\cite{Sc66}. The problem of finding optimum transmission schemes for noisy feedback has been an important open problem. In this problem, both the sender and the receiver receive different observations whose domain increase exponentially in time, and the set of possible strategies grow double exponential in time. Because of asymmetry of information and lack of any common information, there is no known (dynamic programming like) methodology that decomposes this problem in time reducing the complexity to linear in time. Despite the lack of a proper mathematical treatment, recently it was shown in~\cite{Jietal19} that a scheme using RNN (recurrent neural networks) improve the current best known scheme by three orders of magnitude. 

In this paper, we present a sequential decomposition framework that provides a concept of state and allows to decompose this problem across time, to find optimal Markovian policies (w.r.t. that state, that are not necessarily globally optimum). By doing so it provides a framework to reduce the time complexity from exponential to linear. To the best of our knowledge, this is the first instance of decentralized stochastic control problem without any common information, that allows sequential decomposition.

We consider policies of the sender such that current transmission is a function of the sender's message and a controlled Markov process that sender updates on observing the feedback. 
The receiver maintains a belief on the message and the controlled Markov process of the sender to decode the message. This belief is updated using the sender's policy function and it does ML decoding on this belief at the last stage to obtain an estimate of the message. Now the receiver's role is absent from the problem formulation.

Equivalently, there is only the sender who observes the message and the noisy feedback, based on which it maintains a controlled Markov process. It has a cost function that is a function of the receiver's belief, which it doesn't observe perfectly. So the sender puts a belief on this state conditioned on its information, which is now a state of the system that the sender perfectly observes. This state is controlled by sender's policy function at time $t$. Based on this, we formulate a dynamic programming in sender's belief as state and its policy function as its action.

In the following, we denote random variables with capital letters $X, Y, Z,...$, their realizations with small letters $x, y, z, ...$, and alphabets with calligraphic letters $\cX, \cY, \cZ,...$. A sequence is denoted with $X^1_{1:t} = (X^1_1,...,X^1_t)$.
We use the notation $\P(x|y)$ to denote $\P(X=x|Y=y)$.
The space of probability distributions (or equivalently probability mass functions) on the finite alphabet $\cX$ is denoted by $\cP(\cX)$.

\vspace*{-0.15cm}
\section{Channel Model}
\label{sec:model}

We consider a point to point (PTP) discrete memoryless channel (DMC) with noisy feedback. The input symbols $W,X$, $Y,Z$ take values in the finite alphabets $\cW,\cX$, $\cY$ and $\cZ$, respectively. %Our model considers noiseless feedback, that is, the presence of the channel output $z_{1:t-1}$ to both encoders with unit delay.

Consider the problem of transmission of messages $W\in \cW=\{1,\ldots,M\}, \; i=1,2$, over the PTP DMC with noisy feedback using fixed length codes of length $n$.
Encoder generates its channel inputs based on its private message $W$ and noisy feedback $Z_{1:t-1}$. Thus
\begin{align}
X_t &= \tilde{f}_t(W,X_{1:t-1},Z_{1:t-1})=f_t(W,Z_{1:t-1}), \quad i=1,2.
\end{align}
The decoder estimates the messages $W$ based on $n$ channel outputs, $Y_{1:n}$ as
\begin{equation}
\hat{W} = g(Y_{1:n}).
\end{equation}
The channel is memoryless in the sense that the current channel output is independent of all the past channel inputs and the channel outputs, i.e.,
\begin{equation}
\P(y_t|x_{1:t}, y_{1:t-1}, z_{1:t-1}) = Q^f(y_t|x_t).
\end{equation}
Finally, after each transmission, the sender receives a noisy feedback of the transmission $Z_t$ as
\begin{equation}
\P(z_t|y_{1:t}) = Q^b(z_t|y_t)
\end{equation}

A fixed-length transmission scheme for the channels $Q^f,Q^b$ is the pair $s=(f,g)$, consisting of the encoding functions
$f$ and decoding function $g$.
The error probability associated with the transmission scheme $s$ is defined as
\begin{equation}
Pe(s) = \P^S(W\neq \hat{W}).
\end{equation}

%A further generalization of these schemes considers randomized encoding functions, i.e.,
%\begin{align}
%\Pi_T \sim f_t(\cdot|W,X_{1:t-1},Z_{1:t-1}), \qquad i=1,2,
%\end{align}
%where $f_t: \cW \times \cX^{t-1}\times \cZ^{t-1}\rightarrow \cP(\cX)$ or even randomized encoding functions with a common randomness (common between the senders and the receiver), i.e.,
%\begin{align}
%\Pi_T = f_t(W,X_{1:t-1},Z_{1:t-1},U_t), \qquad i=1,2,
%\end{align}
%where $\P(u_t|u_{1:t-1},x^1_{1:t-1},x^2_{1:t-1},z_{1:t-1})=\P(u_t)=u(u_t)$, with $u(\cdot)$ the uniform distribution over $[0,1]$. In this case, the decoder is of the form $(\hat{W}^1,\hat{W}^2) = g(Z_{1:n},U_{1:n})$.
%
%For simplicity of exposition we only consider fixed-length schemes, although the model can be generalized to
%variable-length schemes and the subsequent structural results are valid in that case as well.

\section{Decentralized control of PTP DMC with noisy feedback}
\label{sec:dsaht}

One may pose the following optimization problem.
Given the alphabets $\cW,\cX$, $\cY$, $\cZ$, the channels $Q^f,Q^b$, the pair $(M_1,M_2)$, and for a fixed length $n$, design the optimal transmission scheme $s=(f,g)$ that minimizes the error probability $P_e(s)$.
\begin{equation}
Pe^* = \min_s Pe(s) \tag{\textbf{P1}}
\end{equation}

%In the following we reformulate the problem \textbf{(P1)} into an equivalent optimization problem.
%%
%Using the ``common agent'' methodology for decentralized dynamic team problems~\cite{NaMaTe13}, we now decompose the encoding process $X_t=f_t(W,Z_{1:t-1})$ into an equivalent two-stage process.
%In the first stage, based on the {common information $Z_{1:t-1}$}, the mapping (or ``partial encoding function'') $\phi_t$,  are generated as $\phi_t=\phi_t[z_{1:t-1},u_{1:t},w]$\footnote{We use square brackets to denote functions with range being function sets, i.e., we use notation $\phi_t=\phi_t[z_{1:t-1},u_{1:t},w]$ because $\phi_t$ is itself a function.} where $\phi_t : \cW \rightarrow \cX$. In the second stage, each of these mappings are evaluated at the {private information of each agent}, producing $X_t=\phi_t(W)$.
%%
%In other words, let $\cE$ be the collection of  all encoding functions $e: \cW \rightarrow \mathcal{X}$.
%In the first stage, the common information given by $Z_{1:t-1}$ is transformed using mappings $\phi_t: \cZ^{t-1}  \rightarrow \cE$ to produce an encoding function $\phi_t$. In the second stage these functions are evaluated at the private message $W$ producing $X_t=\phi_t(W)=\phi_t[z_{1:t-1}](W)$.

For any pair of encoding functions, the optimal decoder is the ML decoder (assuming equally likely hypotheses), denoted by $g_{ML}$.
Thus we have reformulated problem (\textbf{P1}) as
\begin{equation}
Pe^* = \min_{\phi} Pe(\phi),  \tag{\textbf{P2}}
\end{equation}
where we have defined $Pe(\phi)$ with a slight abuse of notation based on the above equivalence between encoding functions $f$ and mappings $\phi$, as well as the use of ML decoding.

In this following, we will provide a sequential decomposition methodology to find optimal policies within the class of policies that satisfy $x_t = \phi_t(w,u_t)$, where $\{u_t, \phi_t(u_t,\cdot)\}$ is a controlled Markov Process such that, for any given function $G^2$,
\eq{
u_{t+1} &:= G^2(u_t,\phi_t(u_t,\cdot),z_{t},w).
}

Let 
\eq{
\pi_t^R(w) &:= P^{\phi}(w|y_{1:t}),\\
 \eta_t^R(u_t,u_{t+1}) &:= P^{\phi}(u_t,u_{t+1}|y_{1:t},w), \\
 \pi_t^S(\pi_{t-1}^R,\eta_{t-1}^R) &:= P^{\phi}(\pi_{t-1}^R,\eta_{t-1}^R|z_{1:t-1},w,u_{1:t},x_{1:t}).
 } 
 Then we can easily show that 
\eq{
\pi_{t+1}^R &= F^1(\pi_t^R,\eta_{t+1}^R,\phi_{t+1},y_{t+1})\\
\eta_{t+1}^R &= F^2(\eta_t^R,\phi_{t+1},y_{t+1})\\
\pi_{t+1}^S &= G^1(\pi_t^S,\phi_{t},z_t,w)
}
where 
\eq{
u_t &= f(\pi_{t+1}^S,w)\\
u_{t+1} &= G^2(u_t,\phi_t(z_t,\cdot),z_{t},w)
}
We also assume that $G^2$ is defined such that there is a one-to-one correspondence between $\pi_t^S$ and $u_t$ so that each $\pi_t^S$ leads to one optimal action corresponding to $u_t$.

The ML decoder can now be expressed based on $\pi^R_n$ as
\begin{equation}
\hat{W} = \arg\max_{w} \Pi^R_n(w),
\end{equation}
and the resulting error probability is
\begin{equation}
P_e(\phi) =\E^{\phi}[ 1- \max_{w} \Pi^R_n(w)]=\E^{\phi}[c_{n+1}(\Pi^R_n)],
\end{equation}
where we defined the terminal cost function as
\begin{equation}\label{eq:terminal}
c_{n+1}(\pi^R_n) = 1- \max_{w} \pi^R_n(w),
\end{equation}
and the expectation is w.r.t. the random variable $\Pi^R_{n}$. We define $c_t(\pi^R_t) = 0 \forall t\neq n+1$.

We now show that $\pi_t^R$ can be updated using Bayes rule in a policy-independent way as

\begin{subequations}\label{eq:pi_update1}
\begin{align}
\pi^R_t(w) &= \P^{\phi}(w|y_{1:t}) \\
 &=\frac{\sum_{u_t}\P^{\phi}(w,u_t,y_t|y_{1:t-1})}{\P^{\phi}(y_t|y_{1:t-1})} \\
 &=\frac{\sum_{u_t}\pi^R_{t-1}(w)\P^{\phi}(u_t,y_t|w,y_{1:t-1})}{\P^{\phi}(y_t|y_{1:t-1})} \\
 &=\frac{\sum_{u_t}\pi^R_{t-1}(w) \eta_{t-1}^R(u_t)Q(y_t|\phi_t(w,u_t))}
        {\sum\limits_{\tw,u_t} \pi^R_{t-1}(\tw) \eta_{t-1}^R(u_t)Q(y_t|\phi_t(\tw,u_t))}.
\end{align}
Thus 
\eq{
\pi_t^R = F^1(\pi_{t-1}^R,\eta_{t-1},\phi_t,y_t)
}
\end{subequations}

\begin{subequations}\label{eq:pi_update2}
\begin{align}
\eta^R_t(u_t,u_{t+1}) &= \P^{\phi}(u_t,u_{t+1}|y_{1:t},w) \\
 &=\frac{\P^{\phi}(u_t,u_{t+1},y_t|y_{1:t-1},w)}{\P^{\phi}(y_t|y_{1:t-1})} \\
 %&=\frac{\eta^R_{t-1}(u_{t})\P^{\phi}(u_t,y_t|w,y_{1:t-1})}{\P^{\phi}(y_t|y_{1:t-1})} \\
 &=\frac{\sum_{z_t}\eta^R_{t-1}(u_{t}) Q(y_{t}|\phi_{t}(w,u_{t}))Q(z_{t}|y_{t})I(u_{t+1}=G^2(u_{t},\phi_{t}(z_t,\cdot),z_{t},w))}
        {\sum_{u_t,z_t}\eta^R_{t-1}(u_{t})Q(y_{t}|\phi_{t}(w,u_{t}))Q(z_{t}|y_{t})I(u_{t+1}=G^2(u_{t},\phi_{t}(z_t,\cdot),z_{t},w))}.
\end{align}
Thus 
\eq{
\eta_t^R = F^2(\eta_{t-1}^R,\phi_t,y_t)
}
\end{subequations}

\begin{subequations}
\label{eq:pi_update3}
\begin{align}
\pi^S_t(\pi_t^R,\eta_t^R) &= \P^{\phi}(\pi_t^R,\eta_t^R|z_{1:t},w,x_{1:t},u_{1:t}) \\
 &=\frac{\sum_{\pi_{t-1}^R,\eta_{t-1}^R,y_t}\P^{\phi}(\pi_{t-1}^R,\eta_{t-1}^R,x_t,y_t,z_t|z_{1:t-1},w,x_{1:t-1})}{\P^{\phi}(y_t|z_{1:t},w,u_{1:t},x_{1:t})} \\
 %&=\frac{\sum_{\pi_{t-1}^R,\eta_{t-1}^R,y_t}\pi^S_{t-1}(\pi_{t-1}^R,\eta_{t-1}^R)\P^{\phi}(u_t,y_t|w,y_{1:t-1})}{\P^{\phi}(y_t|y_{1:t-1})} \\
 &=\frac{\sum_{\pi_{t-1}^R,\eta_{t-1}^R,y_t}\pi_{t-1}^S(\pi_{t-1}^R,\eta_{t-1}^R)Q^f(y_t|\phi_t(w,u_t))I(\pi_{t}^R,\eta_{t}^R =F(\pi_{t-1}^R,\eta_{t-1}^R,\phi_t,y_t)) Q^b(z_t|y_t)}
        {\sum_{\pi_{t-1}^R,\eta_{t-1}^R,y_t}\pi_{t-1}^S(\pi_{t-1}^R,\eta_{t-1}^R)Q^f(y_t|\phi_t(w,u_t))I(\pi_{t}^R,\eta_{t}^R =F(\pi_{t-1}^R,\eta_{t-1}^R,\phi_t,y_t)) Q^b(z_t|y_t)}.
\end{align}
Thus 
\eq{
\pi_t^S = G^1(\pi_{t-1}^S,\phi_{t},z_t,u_t,w)
}
\end{subequations}
We summarize the above result into the following lemma.
\begin{lemma}
The posterior belief $\pi^R_t$ on the message $W$, $\eta_t^R$ on $u_t,u_{t+1}$ and $\pi_t^S$ on $\pi_t^R$ can be updated in a policy-independent (i.e., $\phi$-independent) way as $\pi^R_t=F^1(\pi^R_{t-1},\phi_t,y_t), \eta_t^R = F^2(\eta_{t-1}^R,\phi_t,y_t),\pi^S_t=G^1(\pi^S_{t-1},\phi_t,z_t,u_t,w)$.
\end{lemma}
\begin{IEEEproof}
The proof is given in~\eqref{eq:pi_update1}--\eqref{eq:pi_update3}.
\end{IEEEproof}
Based on the above, we present a dynamic program for the sender as follows.

For all $\pi_t^S, w$,
\begin{subequations}\label{eq:bdp}
\begin{align}
V^S_{n+1}(\pi^S_n,w) &= c_{n+1}(\pi^S_n) \\
V^S_t(\pi^S_{t-1},u_t,w) &= c_{n}(\pi^S_{n-1}) + \min_{\phi_t}  \E [ V^S_{t+1}(G^1(\pi^S_{t-1},\phi_t,Z_t,w)) | \pi^S_{t-1},\phi_t,u_t ,w] \label{eq:Vdef}\\
               &= c_{n}(\pi^S_{n-1}) + \min_{\phi_t} \sum_{z_t,w} Q^f(y_t|\phi_t(w,u_t))Q^b(z_t|y_t) V_{t+1}^S(G^1(\pi^S_{t-1},\phi_t,z_t,u_t,w)).
\end{align}
\end{subequations}

All the above results can be summarized in the following theorem
\begin{theorem}\label{Thm:Main}
For the optimization problem~\textbf{(P1)}, one can find optimal Markovian policies of the kind $x_t = \phi_t(w,z_t)$ with state at time $t$, $\pi^S_{t-1}$; action $\phi_t$; zero instantaneous costs $c_t(\pi^S_{t-1},\phi_t)=0$ for $t=1,\ldots,n$; and terminal cost $c_{n+1}(\pi^S_n)$ given in~\eqref{eq:terminal}. Consequently, the optimal encoders are of the form $x_t=\phi_t(w,z_t)$, where $\phi$ can be found through backward dynamic programming as in~\eqref{eq:bdp}.
\end{theorem}
\begin{IEEEproof}
Please see Appendix~A.
\end{IEEEproof}

%Question: Is there a capacity expression for this?
\subsection{Conjecture}
In this section, we conjecture that there exists a PMS~\cite{ShFe08} like scheme such that
\begin{align}
\xi_t &:= \pi_t^R(W)\\
x_t &= F_X^{-1}(\xi_t) \text{ with probability } \P(\xi_t|\pi_t^S)
\end{align}
that achieves capacity, where $F_X$ is the capacity achieving distribution.

We note that in case of noiseless feedback i.e. when $z_t=y_t$ w.p. 1, then $\pi_t^S(\pi) = \delta_{\pi_t^R}(\pi),$ where $\pi_t^R$ is the actual belief of the receiver, i.e. sender perfectly observes the receiver's belief and the above conjectured scheme boils down to the PMS scheme.

\section{Conclusion}
In this paper, we considered point to point discrete memoryless channel with noisy feedback. This falls into the purview of decentralized stochastic control where both the controllers have no common information. Thus, the standard tools in the literature do not apply directly. In this paper, we show that despite there being no common information, there does exist a dynamic programming methodology to compute optimum Markovian policies of the senders involving a belief on belief state. We also conjecture a transmitting scheme that is inspired by the PMS scheme that minimizes probability of error. We ask if based on the above framework, it is possible to design (possibly suboptimal) schemes that are easy to implement.

\section{Acknowledgement}
The author would sincerely like to thank Achilleas Anastasopoulos for valuable comments and (noiseless) feedback.
\appendices
\section{(Proof of Theorem~\ref{Thm:Main})}
\label{app:Theorem_Team}
\begin{IEEEproof}
we will prove $\forall t$
\eq{
&\E^{\psi_{1:t-1}\phi_{t:T},\pi^S_t} \left\{ \sum_{n=t}^T c_n(\Pi^R_n) \big\lvert z_{1:t-1},u_{1:t},w\right\} \leq 
\E^{\psi_{1:t-1}\psi_{t:T},\pi^S_t} \left\{ \sum_{n=t}^T c_n(\Pi^R_n) \big\lvert z_{1:t-1},u_{1:t},w\right\}. \label{eq:prop}
}

The above theorem implies for $t=1$ that 
\eq{
&\E^{\phi_{1:T},\pi^S_t} \left\{ \sum_{n=1}^T c_n(\Pi^R_n) \big\lvert w\right\} \leq 
\E^{\psi_{t1:T},\pi^S_t} \left\{ \sum_{n=1}^T c_n(\Pi^R_n) \big\lvert w\right\}. \label{eq:prop}
}

We prove (\ref{eq:prop}) using induction and from results in Lemma~\ref{lemma:2} and \ref{lemma:1} proved in Appendix~\ref{app:lemmas}. 
\seq{
With slight abuse of notation, let $\pi^S_t$ be a belief function that maps sender's information $z_{1:t-1},u_{1:t},w$ to a belief on $\pi_t^R$ and is consistent (using Bayes' rule) with given past history of policies.
For base case at $t=T$, $\forall z_{1:T-1}\in \mathcal{H}_{T}^c, \psi,$ where $ \pi^S_T[z_{1:T-1},u_{1:T},w](\pi^R_T) := \P^{\psi_{1:T-1}}(\pi^R_T|z_{1:T-1},u_{1:T},x_{1:T-1},w), \forall \pi_T^S$,
\eq{
&\E^{ \psi_{1:T-1}\phi_{T},\pi^S_T}\left\{  c_T(\Pi^R_T) \big\lvert z_{1:t=T-1},u_{1:T},w\right\}\nn\\
&=V_T(\pi^S_T[z_{1:T-1},u_{1:T},w],w)  \label{eq:T2a}\\
&\leq \E^{ \psi_{1:T},\pi^S_T} \left\{ c_T(\Pi^R_T) \big\lvert z_{1:T-1},u_{1:T},w\right\}  \label{eq:T2}
}
}
where (\ref{eq:T2a}) follows from Lemma~\ref{lemma:1} and (\ref{eq:T2}) follows from Lemma~\ref{lemma:2} in Appendix~\ref{app:lemmas}.

From induction hypothesis, for $t+1$, $\forall  z_{1:t} , \psi, $ where $\forall (\pi^R_{t+1}), \pi^S_{t+1} [z_{1:t},u_{1:t+1},w](\pi^R_{t+1}) := \P^{\psi_{1:t}}(\pi^R_{t+1}|z_{1:t},u_{1:t+1},w)$,
\seq{
\eq{
 & \E^{ \psi_{1:t}\phi_{t+1:T},\pi^S_{t+1}} \left\{ \sum_{n=t+1}^T c_n(\Pi^R_n) \big\lvert  z_{1:t-1},u_{1:t},w\right\} \leq\nn \\
 &\E^{\psi_{1:T},\pi^S_{t+1}} \left\{ \sum_{n=t+1}^T c_n(\Pi^R_n) \big\lvert z_{1:t-1},u_{1:t},w\right\}. \label{eq:PropIndHyp}
}
}
\seq{
Then $\forall z_{1:t-1}, \psi, $ where $\forall (\pi_t^R), \pi^S_t [z_{1:t-1},u_{1:t},w](\pi_t^R) := \P^{\psi_{1:t-1}}(\pi_t^R|z_{1:t-1},u_{1:t},w)$, we have
\eq{
&\E^{ \psi_{1:t-1}\phi_{t:T},\pi^S_t} \left\{ \sum_{n=t}^T c_n(\Pi^R_n) \big\lvert z_{1:t-1},u_{1:t},w\right\} \nonumber \\
&= V_t(\pi^S_{t-1}[z_{1:t-1},u_{1:t},w],u_t,w)\label{eq:T1}\\
&\leq \E^{ \psi_{1:t},\pi^S_t} \left\{ c_t(\Pi_t^R) + V_{t+1} (\bar{\Pi}_{t+1}^S[z_{1:t-1},Z_t,u_{1:t},U_{t+1},w],w) \big\lvert z_{1:t-1},u_{1:t},w\right\}  \label{eq:T3}\\
&= \E^{ \psi_{1:t},\pi^S_t} \left\{ c_t(\Pi_t^R) +  \E^{ \psi_{1:t}\phi_{t+1:T},\bar{\Pi}_{t+1}^S} \left\{ \sum_{n=t+1}^T c_n(\Pi^R_n)  \big\lvert z_{1:t-1},Z_t,u_{1:t},U_{t+1},w\right\}   \big\vert z_{1:t-1},u_{1:t},w\right\}  \label{eq:T3a}\\
&\leq \E^{ \psi_{1:t},\pi^S_t} \left\{ c_t(\Pi_t^R) +  \E^{ \psi_{1:t}\psi_{t+1:T},\bar{\Pi}_{t+1}^S} \left\{ \sum_{n=t+1}^T c_n(\Pi^R_n)  \big\lvert z_{1:t-1},Z_t,u_{1:t},U_{t+1},w\right\}   \big\vert z_{1:t-1},u_{1:t},w\right\}  \label{eq:T3b}\\
%
%&= \E^{ \psi_{1:t},\pi^S_t} \left\{ c_T(\Pi_T,A_t) + \E^{ \psi_{1:t}\psi_{t+1:T},\pi^S_t} \left\{ \sum_{n=t+1}^T c_n(\Pi^R_n) \big\lvert z_{1:t-1},Z_{t},w\right\} \big\vert z_{1:T-1},u_{1:T},x_{1:T},w\right\}  \label{eq:T4} 
%
%&= \E^{\thetz_{1:t-1}\psi_t} \left\{ c_T(\Pi_T,A_t)+  \E^{\thetz_{1:t-1}\psi_{t:T}}\left\{ \sum_{n=t+1}^T c_n(\Pi^R_n) \big\lvert z_{1:t-1},Z_{t},w\right\} \big\vert z_{1:t-1},u_{1:t},w\right\}  \label{eq:T5}\\
%
&=\E^{\psi_{1:T},\pi^S_t} \left\{ \sum_{n=t}^T c_n(\Pi^R_n) \big\lvert z_{1:t-1},u_{1:t},w\right\}  \label{eq:T6},
}
}
where $\forall \pi_{t+1}^R, \bar{\Pi}_{t+1}^S[z_{1:t-1},Z_t,u_{1:t},U_{t+1},w](\pi_{t+1}^R):=\P^{\psi_{1:t}}(\pi^R_{t+1}|z_{1:t-1},Z_t,u_{1:t},U_{t+1},w)$, (\ref{eq:T1}) follows from Lemma~\ref{lemma:1}, (\ref{eq:T3}) follows from Lemma~\ref{lemma:2}, (\ref{eq:T3a}) follows from Lemma~\ref{lemma:1}, (\ref{eq:T3b}) follows from induction hypothesis in (\ref{eq:PropIndHyp}). This completes the induction step.
\end{IEEEproof}

\section{}
\label{app:lemmas}
\begin{lemma}
\label{lemma:2}
$\forall t\in \mathcal{T}, z_{1:t-1}, \psi$ where $\forall (\pi_t^R), \pi^S_t [z_{1:t-1},u_{1:t},w](\pi_t^R) := \P^{\psi_{1:t-1}}(\pi_t^R|z_{1:t-1},u_{1:t},w)$
\eq{
&V_t(\pi^S_{t-1}[z_{1:t-1},u_{1:t},w],u_t,w) \leq \E^{ \psi_{1:t},\pi^S_t} \left\{ c_t(\Pi_t^R) + V_{t+1}(\bar{\Pi}_{t+1}^S[z_{1:t-1},Z_t,u_{1:t},U_{t+1},w],W) \big\lvert  z_{1:t-1},u_{1:t},w\right\}.\label{eq:lemma2}
}
where $\forall \pi_{t+1}^R, \bar{\Pi}_{t+1}^S[z_{1:t-1},Z_t,u_{1:t},U_{t+1},w](\pi_{t+1}^R):=\P^{\psi_{1:t}}(\pi^R_{t+1}|z_{1:t-1},Z_t,u_{1:t},U_{t+1},w)$.
\end{lemma}

\begin{IEEEproof}
We prove this lemma by contradiction.

Suppose the claim is not true. This implies $\exists t,\hat{\psi}_{1:t}, \hz_{1:t-1},\hu_{1:t},\hw,$ and $\hu_{1:t},\hx_{1:t}$ generated from $\hat{\psi}_{1:t},\hz_{1:t-1}$ such that $\forall (\pi_t^R), \hat{\pi}^S_{t-1}[\hz_{1:t-1},\hu_{1:t},w](\pi_t^R) := \P^{\hat{\psi}_{1:t-1}}(\pi_t^R|\hz_{1:t-1},\hu_{1:t},\hx_{1:t-1},\hw)$,\\
$\forall \pi_{t+1}^R, \hat{\Pi}_{t+1}^S[\hz_{1:t-1},Z_t,\hu_{1:t+1},\hw](\pi_{t+1}^R):=\P^{\hat{\psi}_{1:t}}(\pi^R_{t+1}|\hz_{1:t-1},Z_t,\hu_{1:t+1},\hw)$, such that 
\eq{
&\E^{\hat{\psi}_{1:t},\hat{\pi}^S_t} \left\{ c_t(\Pi_t^R) +  V_{t+1} (\hat{\Pi}_{t+1}^S[\hz_{1:t-1},Z_t,\hu_{1:t+1},\hw],W) \big\lvert \hz_{1:t-1},\hu_{1:t},w\right\} \nn \\
&< V_t(\hat{\pi}^S_{t-1}[\hz_{1:t-1},\hu_{1:t},\hw],\hu_t,\hw).\label{eq:E8}
}
%\hwhere ${\xi}_t^{i,*}(\pi_t^R) := P^{\phi}(\pi_t^R| \hz_{1:t-1}, \hat{\hw}_{1:t})$ and $\pi^S_t^*[\hz_{1:t-1},\hu_{1:t},\hw](\pi_t^R,) = P^{\phi}(\pi_t^R,|\hz_{1:t-1})$.
We will show that this contradicts the definition of  $V_t$ in (\ref{eq:Vdef}).

Construct $\hat{\phi}_t$ such that $\forall x_t,w$, $\hat{\phi}_t(x_t|w,\hu_t) = \hat{\psi}_t(x_t|\hz_{1:t-1},\hu_{1:t},w)$, where $\hu_{2:t} = g^2(u_1,\hat{\psi}_{1:t-1},\hz_{1:t-1})$.

Then for $\hz_{1:t-1},\hu_{1:t},\hw$, we have
\seq{
\eq{
&V_t(\hat{\pi}^S_{t-1}[\hz_{1:t-1},\hu_{1:t},\hw],\hu_t,\hw) \nn \\
&> \E^{\hat{\psi}_{1:t},\hat{\pi}^S_t} \left\{ c_t(\Pi_t^R) +  V_{t+1} (\hat{\Pi}_{t+1}^S[\hz_{1:t-1},Z_t,\hu_{1:t+1},\hw],W) \big\lvert\hz_{1:t-1},\hu_{1:t},w\right\}   \label{eq:E10}\\
&=\sum_{\pi_t^R,x_t,y_{t},u_t }\left[ c_t(\pi_t^R) +   V_{t+1} (\hat{\Pi}_{t+1}^S[\hz_{1:t-1},Z_t,\hu_{1:t+1},\hw],U_{t+1},\hw )\right]\hat{\pi}^S_{t-1}[\hz_{1:t-1},\hu_{1:t},\hw](\pi_t^R)\nn\\
& \hat{\psi}_t(x_t|\hz_{1:t-1},\hu_{1:t},\hw)Q^f(y_t|x_t)Q^r(z_t|y_t)I(u_{t+1} = G^2(\hu_t,\hat{\psi}_t(\hz_{1:t-1},\cdot),z_t)) \label{eq:E11}\\
&=\sum_{\pi_t^R,x_t,y_t}\left[ c_t(\pi_t^R) +   V_{t+1}( G^1(\hat{\pi}^S_{t-1}[\hz_{1:t-1},\hu_{1:t},\hw],\hat{\phi}_t,z_t,u_{t+1},\hw),\hw ) \right]\hat{\pi}^S_{t-1}[\hz_{1:t-1},\hu_{1:t},\hw](\pi_t^R)\nn\\
& \hat{\phi}_t(x_t|\hw)Q^f(y_t|x_t)Q^r(z_t|y_t)I(u_{t+1} = G^2(\hu_t,\hat{\phi}_t(\hu_t,\cdot),z_t))  \label{eq:E11b}\\
%
%&=\E^{\hat{\phi}_t(\cdot|{\xi}_t) \bet\Pi_T^{*,-i},\,\pi^S_t^*[\hz_{1:t-1},\hu_{1:t},\hw],{\xi}_t^{i,*}[\hz_{1:t-1},  \hat{\hw}_{1:t},\hw]} \left\{ c_T(\Pi_T,A_t) + V_{t+1} (\pi^S_{t+1}^*[\hz_{1:t-1},Z_t]^{i,*}[\hz_{1:t-1},\hat{\hw}_{1:t},W_{t+1},\hw] ) \big\lvert {\xi}_t^{i,*}[ \hz_{1:t-1},  \hat{\hw}_{1:t},\hw],\hw\right\} \\
%
&\geq \min_{\phi_t} \E^{\phi_t,\hat{\pi}^S_t } \left\{ c_t(\Pi_t^R) +  V_{t+1} (G^1(\hat{\pi}^S_{t-1}[\hz_{1:t-1},\hu_{1:t},\hw],{\phi}_t,Z_t,U_{t+1},W),W ) \big\lvert  \hz_{1:t-1},\hu_{1:t},\hw\right\}, \\
&= V_t(\hat{\pi}^S_{t-1}[\hz_{1:t-1},\hu_{1:t},\hw],\hu_t,\hw) \label{eq:E12}
}
where (\ref{eq:E10}) follows from (\ref{eq:E8}), \eqref{eq:E11b} follows from the definition of $\hat{\phi}_t$ and Lemma~\ref{lemma:4a}, and (\ref{eq:E12}) follows from the definition of $V_t$ in (\ref{eq:Vdef}). This leads to contradiction.
}
\end{IEEEproof}

\begin{lemma}
\label{lemma:4a} Suppose for some $\hz_{1:t-1},\hu_{1:t},\hw$, and $\forall (\pi_t^R,\eta_t^R), \\
\pi^S_t [\hz_{1:t-1},\hu_{1:t},\hw](\pi_t^R,\eta_t^R) := P^{{\psi}_{1:t-1}}(\pi_t^R,\eta_t^R|\hz_{1:t-1},\hu_{1:t},\hw)$. Then $\forall (\pi^R_{t+1},\eta_{t+1}^R)$
\eq{
\bar{\pi}_{t+1}^S[\hz_{1:t-1},z_t,\hu_{1:t+1},\hw] (\pi^R_{t+1},\eta_{t+1}^R):= P^{{\psi}_{1:t}}(\pi^R_{t+1},\eta_{t+1}^R|\hz_{1:t-1},z_t,\hu_{1:t+1},\hw) \\=  G^1(\pi^S_{t-1}[\hz_{1:t-1},\hu_{1:t},\hw],{\phi}_t,z_t,\hu_t,w)(\pi^R_{t+1},\eta_{t+1}^R)
}
for ${\phi}_t(x_t|\hw,\hu_t) = {\psi}_t(x_t|\hz_{1:t-1},\hu_{1:t+1},\hx_{1:t},\hw) $, where $\hu_{1:t} = G^2({\psi}_{1:t-1}, \hz_{1:t-1},w)$.
\end{lemma}
\begin{IEEEproof}
		\seq{
	\eq{
	&P^{{\psi_{1:t}}}(\pi^R_{t+1},\eta_{t+1}^R|\hz_{1:t-1},z_t,\hu_{1:t+1},\hw) \\
	&=\frac{\sum_{\pi^R_{t},\eta_t^R,x_t,y_t}P^{{\psi_{1:t}}}(\pi^R_{t},\eta_t^R,x_t,y_t,z_t,\pi^R_{t+1}|\hz_{1:t-1},\hu_{1:t},\hw)}{\sum_{\pi^R_{t},\eta_t^R,x_t,y_t}P^{{\psi_{1:t}}}(\pi^R_{t},\eta_t^R,x_t,y_t,z_t|\hz_{1:t-1},\hu_{1:t},\hw)}\\
	&=\frac{\sum_{\pi^R_{t},\eta_t^R,x_t,y_t}\pi_t^S[\hz_{1:t-1},\hu_{1:t},\hw](\pi^R_{t},\eta_t^R)\psi(x_t|\hz_{1:t-1},\hu_{1:t},\hw)Q^f(y_t|x_t)Q^b(z_t|y_t)I(\pi_{t+1}^R =F(\pi^R_{t},\eta_t^R,\phi_t,y_t))}{\sum_{\pi^R_{t},\eta_t^R,x_t,y_t}\pi_t^S[\hz_{1:t-1},\hu_{1:t},\hw](\pi^R_{t},\eta_t^R)\psi(x_t|\hz_{1:t-1},\hu_{1:t},\hw)Q^f(y_t|x_t)Q^b(z_t|y_t)}\\
	&=\frac{\sum_{\pi^R_{t},\eta_t^R,x_t,y_t}\pi_t^S[\hz_{1:t-1},\hu_{1:t},\hw](\pi^R_{t},\eta_t^R)\phi_t(x_t|\hw,\hu_t)Q^f(y_t|x_t)Q^b(z_t|y_t)I(\pi_{t+1}^R,\eta_{t+1}^R =F(\pi^R_{t},\eta_t^R,\phi_t,y_t))}{\sum_{\pi^R_{t},\eta_t^R,x_t,y_t}\pi_t^S[\hz_{1:t-1},\hu_{1:t},\hw](\pi^R_{t},\eta_t^R)\phi_t(x_t|\hw,\hu_t)Q^f(y_t|x_t)Q^b(z_t|y_t)}\\
	%&=  \sum_{, \pi_t^R, x_{t+1},\hw_{t+1}}P^{{\psi}}(,  \pi_t^R,x_{t+1},\hw_{t+1}, \xi_{t+1}|z_{1:t},\gammz_{1:t}) \\
	&=G^1(\pi^S_{t-1}[\hz_{1:t-1},\hu_{1:t},\hw], {\phi}_t,z_t,\hu_t,w)(\pi^R_{t+1},\eta_{t+1}^R)
	%
%	&= \frac{\sum_{\substack{, \pi_t^R,x_{t+1},w_{t+1}}}  \pi^S_t(\pi_t^R,)\prod_{i=1}^N\phi_t(\Pi_T|)Q^\pi_t^R(x_{t+1}|\pi_t^R,\Pi_T)
%	P^{i,w}(w_{t+1}|x_{t+1},\Pi_T) I_{G_t(,w_{t+1},\Pi_T, \pi^S_t, \phi_t^{-i})}(\xi_{t+1})}{\sum_{} \pi^S_t(\pi_t^R, )\prod_{i=1}^N \phi_t(\Pi_T|)}\\
	%
	%
	}
%	Thus we have,
%\eq{
%P^{{\psi_{1:t}}}(\pi^R_{t+1}|\hz_{1:t-1},z_t,w)	=  G(\pi^S_{t-1}[\hz_{1:t-1},\hu_{1:t},\hw], {\phi}_t,z_t)(\pi^R_{t+1}) ~\label{eq:F_update}
%	}
	}
	\end{IEEEproof}

\begin{lemma}
\label{lemma:1}
$\forall t\in \mathcal{T},\psi, z_{1:t-1},$ where $\forall (\pi_t^R,\eta_t^R),$\\ $\pi^S_t [z_{1:t-1},u_{1:t},w](\pi_t^R,\eta_t^R) := P^{\psi_{1:t-1}}(\pi_t^R,\eta_t^R|z_{1:t-1},u_{1:t},w)$,
\eq{
V_t(\pi^S_{t-1}[z_{1:t-1},u_{1:t},w],u_t,w) 
&= \E^{ \psi_{1:t-1}\phi_{t:T},\pi^S_t} \left\{ \sum_{n=t}^T c_n(\Pi^R_n,\eta_t^R) \big\lvert  z_{1:t-1},u_{1:t},w\right\} 
}
%where $\pi^S_t$ is such that $\pi^S_t(\pi_t^R,) = P^{\thetz_{1:t-1}}(\pi_t^R,|z_{1:t-1})$.
\end{lemma}

\begin{IEEEproof}
%This lemma is in the same spirit as the following statement: ``For a controlled Markov process, if Markov policies are played, then the resulting process is a Markov process, where reward-to-go at any time can be denoted by a function of the current state."

\seq{
We prove the lemma by induction. For $t=T ,\forall \pi_T^R,$ \\$\pi^S_T [z_{1:T-1},u_{1:T},w](\pi_T^R,\eta_T^R) := P^{\psi_{1:T-1}}(\pi_T^R,\eta_T^R|z_{1:T-1},u_{1:T},x_{1:T-1},w)$, 
\eq{
 &\E^{\psi_{1:T-1}\phi_{T},\pi^S_t} \left\{  c_T(\Pi^R_T) \big\lvert z_{1:T-1},u_{1:T},x_{1:T},w\right\}\nonumber \\
 %&= \sum_{\Pi_T,a_T,} c_T(\Pi_T,a_T) P^{\phi_{1:T}}(\Pi_T,a_T,\big\lvert z_{1:T-1})\\
 &= \sum_{\Pi_T^R} c_T(\pi^R_T)\pi^S_{T}[z_{1:T-1},u_{1:T},w](\pi^R_T)\label{eq:C0}\\
 &=V_T(\pi^S_T[z_{1:T-1},u_{1:T},w],u_t,w) \label{eq:C1},
}
}
where (\ref{eq:C1}) follows from the definition of $V_t$ in (\ref{eq:Vdef}).

Suppose the claim is true for $t+1$, i.e., $\forall t\in \mathcal{T}, z_{1:t},\forall (\pi^R_{t+1}), \pi^S_{t+1} [z_{1:t},u_{1:t+1},x_{1:t},w](\pi^R_{t+1}) := \P^{\psi_{1:t}}(\pi^R_{t+1},\eta_{t+1}^R|z_{1:t-1},u_{1:t},w)$
\eq{
&V_{t+1}(\pi^S_{t+1}[z_{1:t},u_{1:t+1},x_{1:t},w],u_t,w) = \E^{\psi_{1:t}\phi_{t+1:T},\pi^S_{t+1}} \left\{ \sum_{n=t+1}^T c_n(\Pi^R_n) \big\lvert z_{1:t-1},u_{1:t},w\right\} \label{eq:CIndHyp}.
}
Then $\forall t\in \mathcal{T}, z_{1:t-1} ,\forall (\pi_t^R), \pi^S_t [z_{1:t-1},u_{1:t},w](\pi_t^R) := \P^{\psi_{1:t-1}}(\pi_t^R,\eta_t^R|z_{1:t-1},u_{1:t},w)$, we have
	\seq{
\eq{
&\E^{\psi_{1:t-1}\phi_{t:T},\pi^S_t} \left\{ \sum_{n=t}^T c_n(\Pi^R_n) \big\lvert  z_{1:t-1},u_{1:t},w\right\} \nonumber \\
&=  \E^{\psi_{1:t-1}\phi_{t:T},\pi^S_t} \left\{c_t(\Pi^R_t) + \E^{\psi_{1:t-1}\phi_{t:T},\pi^S_t} \left\{ \sum_{n=t+1}^T c_n(\Pi^R_n,\eta_n^R)\big\lvert z_{1:t-1}, Z_t,w\right\} \big\lvert z_{1:t-1},u_{1:t},w\right\} \label{eq:C2}\\
&=  \E^{\psi_{1:t-1}\phi_{t:T},\pi^S_t} \left\{c_t(\Pi_t^R) +  \E^{\psi_{1:t-1}\phi_t\phi_{t+1:T}, G^1(\pi^S_t[\cdot],\phi_t,\cdot,w)} \left\{ \sum_{n=t+1}^T c_n(\Pi^R_n)\big\lvert z_{1:t-1},Z_{t},w\right\} \big\lvert z_{1:t-1},u_{1:t},w\right\} \label{eq:C3}\\
%\\
%
&=  \E^{\psi_{1:t-1}\phi_{t:T},\pi^S_t} \left\{c_t(\Pi_t^R) + V_{t+1}(G^1(\pi^S_{t-1}[z_{1:t-1},u_{1:t},w],\phi_t,Z_t,u_t,w),U_{t+1},w) \big\lvert  z_{1:t-1},u_{1:t},w \right\} \label{eq:C4}\\
%
%&=\sum_{\pi_t^R,\Pi_T,}\P^{\phi_{1:T}}(\pi_t^R,\Pi_T,\big\lvert  z_{1:t-1})\left\{c_T(\pi_t^R,\Pi_T) + V_{t+1}(G(\pi^S_t,\phi_t[\pi^S_t],\Pi_T))  ,w\right\}\\
%
&=\sum_{\pi_t^R,z_t,u_{t+1}}\pi^S_{t-1}[z_{1:t-1},u_{1:t},w](\pi_t^R)\phi_t[\pi^S_{t-1}[z_{1:t-1},u_{1:t},w]](x_t|w)Q^f(y_t|x_t)Q^b(z_t|y_t)I(u_{t+1} = G^2(u_t,\phi_t(u_t,\cdot),z_t))\nn\\
&\left\{c_t(\pi_t^R) + V_{t+1}(G^1(\pi^S_{t-1}[z_{1:t-1},u_{1:t},w],\phi_t,z_t,u_t,w),u_{t+1},w) ,w\right\}\\
%
%&=  \E^{\phi_{1:T}} \left\{c_T(\Pi_T,A_t) +  V_{t+1}(\pi^S_{t+1})\big\lvert  z_{1:t-1},u_{1:t},w\right\} \label{eq:C5}\\
%
&=V_t(\pi^S_{t-1}[z_{1:t-1},u_{1:t},w],u_t,w) \label{eq:C6}
}
}
where (\ref{eq:C4}) follows from the induction hypothesis in (\ref{eq:CIndHyp}), and (\ref{eq:C6}) follows from the definition of $V_t$ in (\ref{eq:Vdef}).
\end{IEEEproof}

\bibliographystyle{IEEEtran}
%\bibliography{\string~/Dropbox/Research/bib/IEEEabrv,\string~/Dropbox/Research/bib/deepanshu,\string~/Dropbox/Research/bib/abhinav}

\end{document}